\documentclass[abstracton]{scrartcl}
\usepackage{epsfig}
\usepackage{amsbsy,amsmath,amsfonts,amssymb}
\usepackage{graphicx}
\usepackage{float}
\usepackage{psfrag}
\usepackage[latin1]{inputenc}
\usepackage{color}
\usepackage{wasysym}
\usepackage{colordvi}
\usepackage[T1]{fontenc}

\def\Journal#1#2#3#4{{#1}~{\bf #2} (#4), #3}

\def\NIMA{{\em Nucl.~Instrum.~Methods}~A}

\def\PLB{{\em Phys.~Lett.}~B}
\def\PR{\em Phys.~Rev.}
\def\PRL{\em Phys.~Rev.~Lett.}
\def\PRD{{\em Phys.~Rev.}~D}

\def\EPJ{{\em Eur.~Phys.~J.}~C}

\newcommand{\subrm}[1]{\mbox{\tiny \rm #1}}

\newcommand{\ks}{K_{\subrm{S}}}

\newcommand{\kspipig}{\ks \to \pi^+ \pi^- \gamma}

\newcommand{\aLam}{\overline{\Lambda}}

\newcommand{\lamppi}{\Lambda \to p \pi^-}

\newcommand{\alphapp}{\alpha_{\Lambda \to p \pi^-}}

\newcommand{\aXi}{\overline{\Xi^0}}

\newcommand{\xilampi}{\Xi^0 \to \Lambda \pi^0}
\newcommand{\axilampi}{\aXi \to \aLam   \pi^0}

\newcommand{\alphalp}{\alpha_{\Xi^0 \to \Lambda \pi^0}}

\newcommand{\xilamgam}{\Xi^0 \to \Lambda \gamma}
\newcommand{\axilamgam}{\overline{\Xi^0} \to \overline{\Lambda} \gamma}
\newcommand{\alphalg}{\alpha_{\Xi^0 \to \Lambda \gamma}}
\newcommand{\alphaalg}{\alpha_{\overline{\Xi^0} \to \overline{\Lambda} \gamma}}
\newcommand{\xisiggam}{\Xi^0 \to \Sigma^0 \gamma}
\newcommand{\axisiggam}{\overline{\Xi^0} \to \overline{\Sigma^0} \gamma}
\newcommand{\alphasg}{\alpha_{\Xi^0 \to \Sigma^0 \gamma}}
\newcommand{\alphaasg}{\alpha_{\overline{\Xi^0} \to \overline{\Sigma^0} \gamma}}
\newcommand{\siglamgam}{\Sigma^0 \to \Lambda \gamma}

\newcommand{\bdm}{\begin{displaymath}}
\newcommand{\edm}{\end{displaymath}}
\newcommand{\be}{\begin{equation}}
\newcommand{\ee}{\end{equation}}

\begin{document}

%
%  Title
%
\centerline{EUROPEAN ORGANIZATION FOR NUCLEAR RESEARCH}
%\vspace*{5mm}
\vspace*{-1mm}
{\flushright{CERN-PH-EP-2010-032\\ 18 May 2010\\}}
%{\flushright{Draft 2 \\ \today\\}}
%
\begin{center}
{\bf \huge New Precise Measurements of the \\*[4mm] $\xilamgam$ and $\xisiggam$ Decay Asymmetries}
\end{center}
\begin{center}
\vspace{2mm}
{\Large The NA48/1 Collaboration}\\*[3mm]
%
%\author{A.~Lai},
%\author{D.~Marras}
%\address{Dipartimento di Fisica dell'Universit\`a and Sezione dell'INFN di Cagliari, I-09100 Cagliari, Italy} 
% 
%
J.R.~Batley,
G.E.~Kalmus$\,$\footnotemark[1],
C.~Lazzeroni$\,$\footnotemark[2],
D.J.~Munday,
M.~Patel$\,$\footnotemark[3],
M.W.~Slater$\,$\footnotemark[2],
S.A.~Wotton \\
{\em \small Cavendish Laboratory, University of Cambridge, Cambridge, CB3 0HE, U.K.$\,$\footnotemark[4]} \\*[2mm]
R.~Arcidiacono$\,$\footnotemark[5],
G.~Bocquet,
A.~Ceccucci,
D.~Cundy$\,$\footnotemark[6],
N.~Doble$\,$\footnotemark[7],
V.~Falaleev,
L.~Gatignon,
A.~Gonidec,
P.~Grafstr\"om,
W.~Kubischta,
I.~Mikulec$\,$\footnotemark[8],
A.~Norton$\,$\footnotemark[10],
B.~Panzer-Steindel,
P.~Rubin$\,$\footnotemark[9],
H.~Wahl$\,$\footnotemark[10] \\
{\em \small CERN, CH-1211 Gen\`eve 23, Switzerland} \\*[2mm]
E.~Goudzovski$\,$\footnotemark[2],
P.~Hristov$\,$\footnotemark[3],
V.~Kekelidze,
L.~Litov,
D.~Madigozhin,
N.~Molokanova,
Yu.~Potrebenikov,
S.~Stoynev,
A.~Zinchenko \\
{\em \small Joint Institute for Nuclear Research, Dubna, Russian Federation} \\*[2mm]
E.~Monnier$\,$\footnotemark[11],
E.~Swallow,
R.~Winston$\,$\footnotemark[12] \\
{\em \small The Enrico Fermi Institute, The University of Chicago, Chicago, IL 60126, U.S.A.} \\*[2mm]
R.~Sacco$\,$\footnotemark[13],
A.~Walker \\
{\em \small Department of Physics and Astronomy, University of Edinburgh, JCMB King's Buildings, Mayfield Road, Edinburgh, EH9~3JZ, U.K.} \\*[2mm]
W.~Baldini,
A.~Gianoli \\
{\em \small INFN Sezione di Ferrara, I-44100 Ferrara, Italy} \\*[2mm]
P.~Dalpiaz,
P.L.~Frabetti$\,$\footnotemark[14],
M.~Martini,
F.~Petrucci,
M.~Savri\'e,
M.~Scarpa \\
{\em \small Dipartimento di Fisica dell'Universit\`a and INFN Sezione di Ferrara, I-44100 Ferrara, Italy} \\*[2mm]
%
%\newpage
%
M.~Calvetti,
G.~Collazuol$\,$\footnotemark[15],
E.~Iacopini,
G.~Ruggiero$\,$\footnotemark[15] \\
{\em \small Dipartimento di Fisica dell'Universit\`a and INFN Sezione di Firenze, I-50125~Firenze, Italy} \\*[2mm]
A.~Bizzeti$\,$\footnotemark[16],
M.~Lenti,
M.~Veltri$\,$\footnotemark[17] \\
{\em \small INFN Sezione di Firenze, I-50125~Firenze, Italy} \\*[2mm]
\renewcommand{\thefootnote}{\fnsymbol{footnote}}
M.~Behler$\,$\footnotemark[1],
\renewcommand{\thefootnote}{\arabic{footnote}}
K.~Eppard,
M.~Eppard,
A.~Hirstius$\,$\footnotemark[3],
K.~Kleinknecht,
U.~Koch,
P.~Marouelli,
L.~Masetti,
U.~Moosbrugger,
C.~Morales Morales,
A.~Peters$\,$\footnotemark[3],
\renewcommand{\thefootnote}{\fnsymbol{footnote}}
R.~Wanke$\,$\footnotemark[1],
\renewcommand{\thefootnote}{\arabic{footnote}}
%R.~Wanke$\,$\thanks{Corresponding Authors.
%          {\it Email addresses:} Matthias.Behler{\@}uni-mainz.de, 
%                                        Rainer.Wanke{\@}uni-mainz.de},
A.~Winhart \\
{\em \small Institut f\"ur Physik, Universit\"at Mainz, D-55099~Mainz, Germany$\,$\footnotemark[18]} \\*[2mm]
A.~Dabrowski$\,$\footnotemark[3],
T.~Fonseca Martin$\,$\footnotemark[3],
M.~Velasco \\
{\em \small Department of Physics and Astronomy, Northwestern Uniwersity, Evanston, IL 60208-3112, U.S.A.} \\*[2mm]
\newpage
P.~Cenci,
P.~Lubrano,
M.~Pepe \\
{\em \small INFN Sezione di Perugia, I-06100 Perugia, Italy} \\*[2mm]
G.~Anzivino,
E.~Imbergamo,
G.~Lamanna$\,$\footnotemark[15],
A.~Michetti,
A.~Nappi,
M.C.~Petrucci,
M.~Piccini,
M.~Valdata \\
{\em \small Dipartimento di Fisica dell'Universit\`a and INFN Sezione di Perugia, I-06100~Perugia, Italy} \\*[2mm]
C.~Cerri,
R.~Fantechi \\
{\em \small INFN Sezione di Pisa, I-56100 Pisa, Italy} \\*[2mm]
F.~Costantini,
L.~Fiorini$\,$\footnotemark[19],
S.~Giudici,
G.~Pierazzini,
M.~Sozzi \\
{\em \small Dipartimento di Fisica, Universit\`a degli Studi di Pisa and INFN Sezione di Pisa, I-56100 Pisa, Italy} \\*[2mm]
I.~Mannelli \\
{\em \small Scuola Normale Superiore and INFN Sezione di Pisa, I-56100 Pisa, Italy} \\*[2mm]
C.~Cheshkov,
J.B.~Cheze,
M.~De Beer,
P.~Debu,
G.~Gouge,
G.~Marel,
E.~Mazzucato,
B.~Peyaud,
B.~Vallage \\
{\em \small DSM/DAPNIA - CEA Saclay, F-91191~Gif-sur-Yvette, France} \\*[2mm]
M.~Holder,
A.~Maier$\,$\footnotemark[3],
M.~Ziolkowski \\
{\em \small Fachbereich Physik, Universit\"at Siegen, D-57068 Siegen, Germany$\,$\footnotemark[20]} \\*[2mm]
%
%\newpage
%
C.~Biino,
N.~Cartiglia,
F.~Marchetto, 
N.~Pastrone \\
{\em \small INFN Sezione di Torino, I-10125~Torino, Italy} \\*[2mm] 
M.~Clemencic$\,$\footnotemark[3],
S.~Goy Lopez$\,$\footnotemark[3],
E.~Menichetti \\
{\em \small Dipartimento di Fisica Sperimentale dell'Universit\`a and INFN Sezione di Torino, I-10125~Torino, Italy} \\*[2mm] 
W.~Wislicki
{\em \small Soltan Institute for Nuclear Studies, Laboratory for High Energy Physics, PL-00-681~Warsaw, Poland$\,$\footnotemark[21]} \\*[2mm]
H.~Dibon,
M.~Jeitler,
M.~Markytan,
G.~Neuhofer,
L.~Widhalm \\
{\em \small \"Osterreichische Akademie der Wissenschaften, Institut  f\"ur Hochenergiephysik, A-1050~Wien, Austria$\,$\footnotemark[22]}
\end{center}

%
%\linenumbers

\vspace*{1mm}

\centerline{\it Accepted by Phys.~Lett.~B}
\vspace*{1mm}

\vspace*{\fill}

\begin{abstract}
The decay asymmetries of the weak radiative Hyperon decays $\xilamgam$ and $\xisiggam$ have been
measured with high precision using data of the NA48/1 experiment at CERN.
From about 52~000 $\xilamgam$ and 15~000 $\xisiggam$ decays, we obtain
for the decay asymmetries 
$\alphalg = -0.704 \pm 0.019_\text{stat} \pm 0.064_\text{syst}$ and
$\alphasg = -0.729 \pm 0.030_\text{stat} \pm 0.076_\text{syst}$, respectively.
These results are in good agreement with previous experiments, but more precise.
% and allow to distinguish between different competing theoretical models.
\end{abstract}

\setcounter{footnote}{0}
%\fnsymbol{footnote}
\renewcommand{\thefootnote}{\fnsymbol{footnote}}
\footnotetext[1]{Corresponding Authors.
          {\it Email:} Matthias.Behler@uni-mainz.de, 
                                        Rainer.Wanke@uni-mainz.de}
\renewcommand{\thefootnote}{\arabic{footnote}}
\setcounter{footnote}{0}
\footnotetext[1]{Present address: Rutherford Appleton Laboratory, Chilton, Didcot, Oxon, OX11~0QX, U.K.}
\footnotetext[2]{Present address: University of Birmingham, Edgbaston, Birmingham, B15 2TT, U.K.}
\footnotetext[3]{Present address: CERN, CH-1211 Gen\`eve 23, Switzerland.}
\footnotetext[4]{Funded by the U.K. Particle Physics and Astronomy Research Council.}
\footnotetext[5]{Present address: Dipartimento di Fisica Sperimentale dell'Universit\`a and INFN Sezione di Torino, \mbox{I-10125} Torino, Italy.}
\footnotetext[6]{Present address: Istituto di Cosmogeofisica del CNR di Torino, I-10133 Torino, Italy.}
\footnotetext[7]{Present address: Dipartimento di Fisica, Universit\`a degli Studi di Pisa and INFN Sezione di Pisa, \mbox{I-56100}~Pisa, Italy.}
\footnotetext[8]{On leave from \"Osterreichische Akademie der Wissenschaften, Institut  f\"ur Hochenergiephysik, \mbox{A-1050}~Wien, Austria.}
\footnotetext[9]{On leave from University of Richmond, Richmond, VA, 23173, U.S.A.; supported in part by the US NSF under award \#0140230.
                 Present address: Department of Physics and Astronomy, George Mason University, Fairfax, VA 22030, U.S.A.}
\footnotetext[10]{Present address: Dipartimento di Fisica dell'Universit\`a and INFN Sezione di Ferrara, I-44100~Ferrara, Italy.}
\footnotetext[11]{Present address: Centre de Physique des Particules de Marseille, IN2P3-CNRS, Universit\'e de la M\'editerran\'ee, Marseille, France.}
\footnotetext[12]{Also at University of California, Merced, U.S.A.}
\footnotetext[13]{Present address: Department of Physics, Queen Mary University, London, E1~4NS, U.K.}
\footnotetext[14]{Present address: Joint Institute for Nuclear Research, Dubna, 141980, Russian Federation.}
\footnotetext[15]{Present address: Scuola Normale Superiore and INFN Sezione di Pisa, I-56100~Pisa, Italy.}
\footnotetext[16]{Also Dipartimento di Fisica dell'Universit\`a di Modena e Reggio Emilia, I-41100 Modena, Italy.}
\footnotetext[17]{Istituto di Fisica, Universit\`a di Urbino, I-61029  Urbino, Italy.}
\footnotetext[18]{Funded by the German Federal Minister for Education and research under contract 7MZ18P(4)-TP2.}
\footnotetext[19]{Present address: Institut de Fisica d'Altes Energies, Facultat Ciencias, Universitat Autonoma de Barcelona, E-08193 Bellaterra, Spain.}
\footnotetext[20]{Funded by the German Federal Minister for Research and Technology (BMBF) under contract 056SI74.}
\footnotetext[21]{Supported by the Committee for Scientific Research grants 5P03B10120, SPUB-M/CERN/P03/DZ210/2000 and SPB/CERN/P03/DZ146/2002.}
\footnotetext[22]{Funded by the Austrian Ministry for Traffic and Research under the contract GZ 616.360/2-IV GZ 616.363/2-VIII, and by the Fonds f\"ur Wissenschaft und Forschung FWF Nr.~P08929-PHY.}
\setcounter{footnote}{0}

%
%  Main Text
%
%---------------------------------------------------------------------------------------------

%\linenumbers

\section{Introduction}

Measurements and theoretical descriptions of weak radiative
hyperon decays have often disagreed.
In 1964 Hara proved that $\Sigma^+$ and $\Xi^-$ decay asymmetries vanish in the SU(3) limit~\cite{bib:hara64}.
Introducing weak breaking of SU(3) symmetry one expects to observe small decay asymmetries~\cite{bib:zenc99}.
In contrast to this, a large negative decay asymmetry in the weak radiative decay $\Sigma^+\to p\gamma$ 
was first measured at Berkeley~\cite{bib:gershwin69} and later confirmed~\cite{bib:pdg08}.
To address this observation, several models were developed which tried to obtain large decay asymmetries
in spite of weak SU(3) breaking. 
One category consists of pole models, which satisfy the Hara theorem by construction,
and approaches based on chiral perturbation theory.
These predict negative decay asymmetries for all weak radiative hyperon decays~\cite{bib:gavela81,bib:borosay99,bib:zenc00}.
Calculations based on vector meson dominance and quark models, using measured data as input, are also 
able to describe the decay asymmetries in a consistent, but not yet fully satisfying picture~\cite{bib:zenc06,bib:dubovik08}.
For discrimination between these different approaches, precise experimental inputs are important.

Measurements of $\Xi^0$ decay asymmetries have been performed by the NA48 experiment,
which has obtained a value of $\alphalg = -0.78 \pm 0.19$ using 730 reconstructed events~\cite{bib:lamgamNA48}, and by the KTeV collaboration, which
found $\alphasg = -0.63 \pm 0.09$ from more than 4000 events~\cite{bib:siggamKTeV}.

In this letter we report on new precise measurements 
of both the $\xilamgam$ and the $\xisiggam$ decay asymmetry,
using a data set at least one order of magnitude larger than those used in previous measurements.

\section{Experimental Apparatus}
\label{sec:experiment}

The NA48/1 experiment took data in 2002 at the CERN SPS. A beam of neutral particles was produced by a $400$~GeV/$c$ proton beam impinging 
on a Be target in 4.8~s long spills repeated every 16.8~s. The proton beam intensity had a mean of $5 \times 10^{10}$ particles per pulse and 
was fairly constant over the duration of the spill.

In the NA48/1 set-up, only the $K_S$ target station of the NA48 double $K_S$/$K_L$ beam line was used~\cite{bib:detector}.
A sweeping magnet deflected charged particles away from the collimators,
which selected a beam of neutral long-lived particles ($K_S$, $K_L$, $\Lambda$, $\Xi^0$, $n$, and $\gamma$).
The defining collimator was located 5.03~m down-stream of the target and had
a circular aperture of 1.8~mm radius, followed by a final collimator, ending 6.23~m down-stream of the target with a radius of 3~mm.
To reduce the number of photons, a 24~mm thick platinum absorber
was placed between the target and the collimators.  
The target and collimator positions were chosen in such a way, that
the beam axis passed through the centre of the electromagnetic calorimeter.
The production angle between the proton beam direction and the axis of the neutral beam was 4.2~mrad.
A right-handed coordinate system was defined with the $z$-axis pointing in direction of the former $K_L$ beam %, the symmetry axis of the experiment,
and the $y$-axis pointing upwards.

The collimator was followed by a 90~m long evacuated tank, with a diameter between 1.92 and 2.4~m and terminated by a $0.3\%$~$X_0$ thick
Kevlar window. The detectors were located down-stream of this region to detect the particles
originating from decays in the tank.
On average, about $1.4 \times 10^4$ $\Xi^0$ hyperons per spill decayed in the fiducial decay volume,
dominantly into the $\Lambda \pi^0$ final state.

%The $\Lambda$ hyperons of the $\xilampi$ decay were detected by measuring the charged particles 
%of the subsequent decay $\Lambda \to p \pi^-$ 
The momenta and positions of charged particles were measured in a magnetic spectrometer.
The spectrometer was housed in a helium gas volume and consisted 
of two drift chambers before and two after a dipole magnet with
vertical magnetic field direction, giving a horizontal transverse momentum kick of 265~MeV/$c$.
Each chamber had four views ($x$, $y$, $u$, $v$) with two sense wire planes each.
The $u$ and $v$ views were rotated by $\pm 45^\circ$ around the $z$ axis with respect to the $x$ and $y$ views.
In the chamber located just down-stream of the magnet, only $x$ and $y$ views were instrumented.
The space points, reconstructed by each chamber, had a resolution of 150~$\mu$m in each projection.
The momentum resolution of the spectrometer was measured to be
$\sigma_p/p = 0.48\% \oplus 0.015\% \times p$, with $p$ in GeV/$c$.
The track time resolution was about 1.5~ns.

Photons were measured with a 27 radiation lengths deep liquid-krypton
electromagnetic calorimeter (LKr). % ~\cite{bib:lkr}.
It was read out longitudinally in about 13500 cells of cross-section $2\times2$~cm$^2$.
The energy resolution was determined to be
$\sigma_E/E = 3.2\%/\sqrt{E} \oplus 9\%/E \oplus 0.42\%$, with $E$ in GeV. %~\cite{bib:eresolution}.
The spatial and time resolutions were better than 1.3~mm and 300~ps, respectively, 
for photons with energies above 20~GeV.

Other detector elements were only used at the trigger level.
An iron-scintillator sandwich hadron calorimeter, 6.7 nuclear interaction lengths thick,
followed down-stream of the LKr. It provided a raw measurement of the energy of hadron showers.
The hadron calorimeter was followed by three planes of scintillation counters, used to detect muons. 
A segmented scintillator hodoscope for charged particles, with a time resolution better than 200~ps for two-track events, was located between
the spectrometer and the LKr calorimeter. 
An additional hodoscope for neutral particles was installed in the LKr calorimeter at a depth of about 9.5 radiation lengths.
Furthermore, seven rings of scintillation counters (AKL) were placed around the decay volume and the helium tank of the spectrometer
to detect activity outside of the detector acceptance.
A more detailed description of the NA48/1 beam-line and detector can be found in~\cite{bib:detector}.

The trigger decision for neutral hyperon decays was based on information from the detector elements described above.
A positive trigger (L1) decision required at least one coincidence between a vertical and a horizontal
scintillator strip of the hodoscope for charged particles, a hit signature in the drift chambers compatible with more than one track, 
no hit in the last two AKL rings,
and the energy sum deposited in the electromagnetic calorimeter larger than 15~GeV or, alternatively, in both calorimeters larger than 30~GeV.
The next trigger level (L2) used information from a preliminary track reconstruction:
at least two oppositely charged tracks were required in the drift chambers,
with an invariant mass being compatible with the nominal $\Lambda$ mass under proton (anti-proton) and $\pi^-$ ($\pi^+$) assumption. 
To suppress $\Lambda$'s originating from the target, a minimum distance of 8~cm in the last drift chamber
of the extrapolated $\Lambda$ line-of-flight from the detector axis was required.
In order to reject $K_S \to \pi^+ \pi^-$ background events, the ratio $p_>/p_<$ between the larger and the smaller track momentum was
required to be larger than 3.5.
This ratio is large for hyperon decays where the proton carries the major fraction of the initial momentum, 
as opposed to background kaon decays where the two charged particles typically have a much lower momentum ratio.
To reject photon conversions, the distance between the two tracks in the first drift chamber was required to be larger than 5~cm. 
Finally, the decay vertex had to be reconstructed within 5~m before and 50~m after the end of the final collimator.
In addition, events with at least four tracks or at least two hits in the muon veto counters were accepted by the L2 trigger.
Because of the high rate, the triggered events used in this analysis were down-scaled by a factor of either 2 or 4
during the run period.

% {\red \em The following was in Sven's paper. Do we need it???}
% The events were selected by a trigger optimised to
% select hadronic events for which the accuracy of the electromagnetic
% calorimeter information was not necessary. For this trigger, showers
% from the LKr were read-out only if their seed energy exceeded 1~GeV
% instead of the usual 100~MeV.
% Therefore a cluster energy correction was applied, which was determined from the
% reconstructed invariant $\pi^0$ mass in $K_L \to \pi^+ \pi^- \pi^0$ events from the same data sample.
% The correction was energy-dependent and ranged between $3\%$ and $0.5\%$ for photon energies between
% 5 and 20~GeV.

A third-level software trigger (L3) used the complete reconstructed detector information.
For selection of radiative hyperon decays at least one $\Lambda \to p \pi^-$ candidate and at least
one LKr energy cluster not associated with any track was required.

\section{Event Selection}

The decays $\xilamgam$, $\xisiggam$, and $\xilampi$ were reconstructed via
the decays $\lamppi$, $\siglamgam$, and $\pi^0 \to \gamma \gamma$.
Each selected event therefore had to have exactly two oppositely charged tracks, consistent with a $\Lambda$ hypothesis, 
and at least one (in case of $\xilamgam$) or two (for $\xisiggam$ and $\xilampi$)
unassociated clusters in the LKr calorimeter. Events with hits in the last two AKL counters, in-time with the average track time, were rejected.

The positive track (proton) had to have a momentum $p_p > 34$~GeV/$c$, 
the negative track (pion) was required to have $p_{\pi^-} > 5$~GeV/$c$.
%The requirement on the proton momentum rejects $\aLamppi$ events, where the $\pi^+$ is misidentified as proton, which arise
%at $p_p < 30$~GeV/$c$ and $p_{\pi^-} > 45$~GeV/$c$. 
The momentum ratio $p_+/p_-$ between positive and negative track had to exceed 3.8.
To reject electrons, the tracks had to have either no associated cluster in the LKr calorimeter or a ratio of cluster energy over momentum less than $0.93$.
Muons were discarded by rejecting tracks with associated in-time hits in the muon counters.

To ensure full detection efficiency in the drift chambers, both tracks were required to have a radial distance
from the detector axis larger than 12.5~cm in the first and the last drift chamber.
The distance between the tracks had to be greater than 10~cm in the first drift chamber 
to reject photon conversions and so-called ghost tracks, which share track segments.
The time difference $\Delta t_\text{tracks}$ between the tracks, measured in the drift chambers, had to be less than 6.5~ns.
The $\Lambda$ decay vertex was defined by the closest approach of the tracks.
The distance of closest approach had to be less than 2.5~cm, and the longitudinal position of the vertex had to be between $-1.5$ and $38$~m,
measured from the end of the final collimator.
The $\Lambda$ line-of-flight was obtained from the sum of the track 4-momenta and the reconstructed decay position.
Events with primary $\Lambda$'s, produced at the target, were removed by requiring the radial position of the extrapolated
$\Lambda$ trajectory in the last drift chamber to be larger than 9~cm.
Finally, the invariant mass of the two tracks under proton and $\pi^-$ assumption 
had to be between 1.1128 and 1.1185~GeV/$c^2$, corresponding to about a $3\sigma$ window around the nominal $\Lambda$ mass.

Due to the different kinematics, 
the minimum reconstructed energy of the photon clusters depended on the decay channel. For $\xilamgam$, it had to be larger than 15~GeV.
For $\xisiggam$ it had to be $> 10$~GeV for the photon from the $\Xi^0$ decay and $> 7$~GeV for the photon from the $\Sigma^0$ decay.
Finally, for $\xilampi$, both photon energies had to be larger than 5~GeV. 
%Each photon cluster was required to have a reconstructed energy greater than $10$, $15$, and $5$~GeV 
%for $\xisiggam$, $\xilamgam$, and $\xilampi$, respectively. 
The clusters had to lie inside the active LKr calorimeter region at a distance of at least 15~cm 
from the detector axis and well inside outer edge.
The closest distance to any dead cell had to be larger than 2~cm.
To reject events with overlapping showers, the cluster width had to be within $3.5 \, \sigma$ of the average cluster width
for the given cluster energy.
Each photon cluster had to have a distance greater than 30~cm from any track impact point at the front surface of the LKr.
Events with additional unassociated clusters, with energies above 2.5~GeV and fulfilling the above geometrical criteria, were rejected.
For $\xisiggam$ and $\xilampi$, the distance between the two clusters had to be larger than 10~cm.
The difference between each cluster time and the mean track time had to be less than 3.5~ns.

An energy centre-of-gravity
\[
(x_\text{cog}, y_\text{cog}) \; = \; \left( \, \frac{\sum_i x_i E_i}{\sum_i E_i}, \, \frac{\sum_i y_i E_i}{\sum_i E_i} \, \right)
\]
at the longitudinal position of the LKr calorimeter
was defined by using the transverse positions $x_i$ and $y_i$ and the energies $E_i$ of the photon clusters and tracks
at the front surface of the LKr calorimeter.
The track energies were computed using their momenta measured in the spectrometer.
The tracks were projected onto the LKr surface from their positions and momenta in the first drift chamber before the spectrometer
magnet.
To suppress badly measured events and possible background from events with lost decay products, the radial distance 
$r_\text{cog} = \sqrt{x_\text{cog}^2 + y_\text{cog}^2}$ of the energy centre-of-gravity to the beam axis was required to be less than 5.5~cm.

The $\Xi^0$ line-of-flight was reconstructed by connecting the target position with the energy centre-of-gravity
in the LKr calorimeter. The closest distance of approach between this line and the extrapolated $\Lambda$ line-of-flight
defined the $\Xi^0$ decay vertex.
Its longitudinal position, which had a resolution of about 2~m for all the decay channels,
was required to be reconstructed not more than 4~m up-stream of the end of the final collimator.
Also, the difference $z_\Lambda - z_{\Xi^0}$ between the longitudinal positions of the reconstructed $\Lambda$ and $\Xi^0$ decay vertices
had to be larger then $-4$~m, taking into account the finite vertex resolutions.

Each photon 4-momentum was reconstructed using the $\Xi^0$ decay vertex and 
the cluster position in the LKr calorimeter.
For $\xilampi$ candidates, the invariant $\gamma \gamma$ mass $m_{\gamma \gamma}$ was required to be between
$125.5$ and $145.5$~MeV/$c^2$, consistent with a $\pi^0$ decay.
For $\xisiggam$, $m_{\gamma \gamma}$ had to be either less than 113~MeV/$c^2$ or greater than
157~MeV/$c^2$, to suppress background from the abundant $\xilampi$ decays.
In addition, the invariant $\Lambda \gamma$ mass had to be between 1.187 and 1.199 GeV/$c^2$, consistent with a $\Sigma^0$ decay.
For $\xilamgam$ candidates, the invariant $\pi^+ \pi^- \gamma$ mass, built under the assumption that both charged particles are pions,
was not allowed to be between 491 and 505~MeV/$c^2$, to reject background from $\kspipig$ decays.

Finally, the $\Xi^0$ candidates were reconstructed from the track and photon 4-momenta.
For all decay channels, the $\Xi^0$ momentum $p$ was required to be above 70 and below 220~GeV/$c$ and the invariant
mass to be between 1.307 and 1.324~GeV/$c^2$ for $\xilamgam$ and between 1.309 and 1.321~GeV/$c^2$ for the other channels.

\begin{figure}[t]
\begin{center}
\epsfig{file=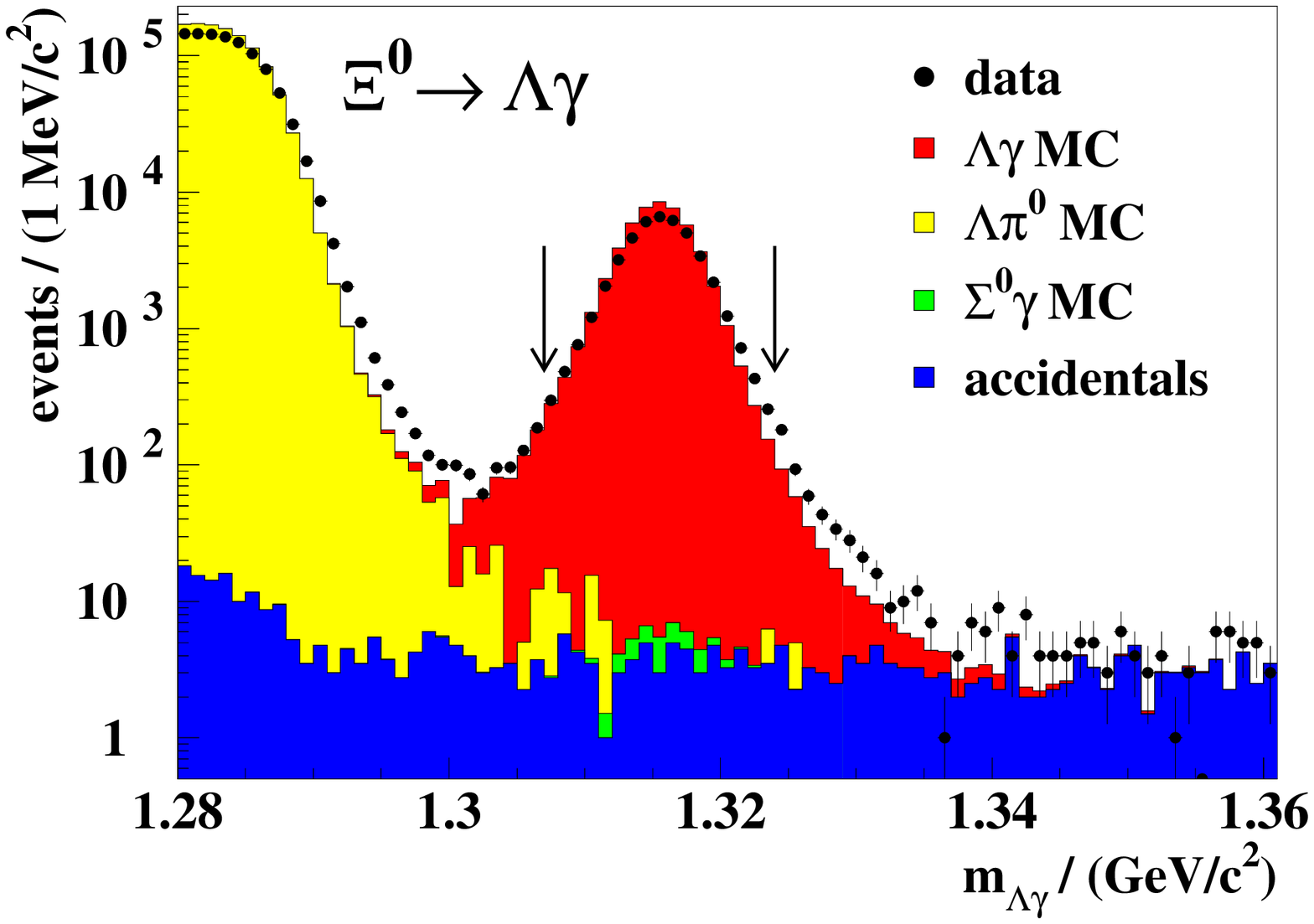,width=0.49\textwidth}
\epsfig{file=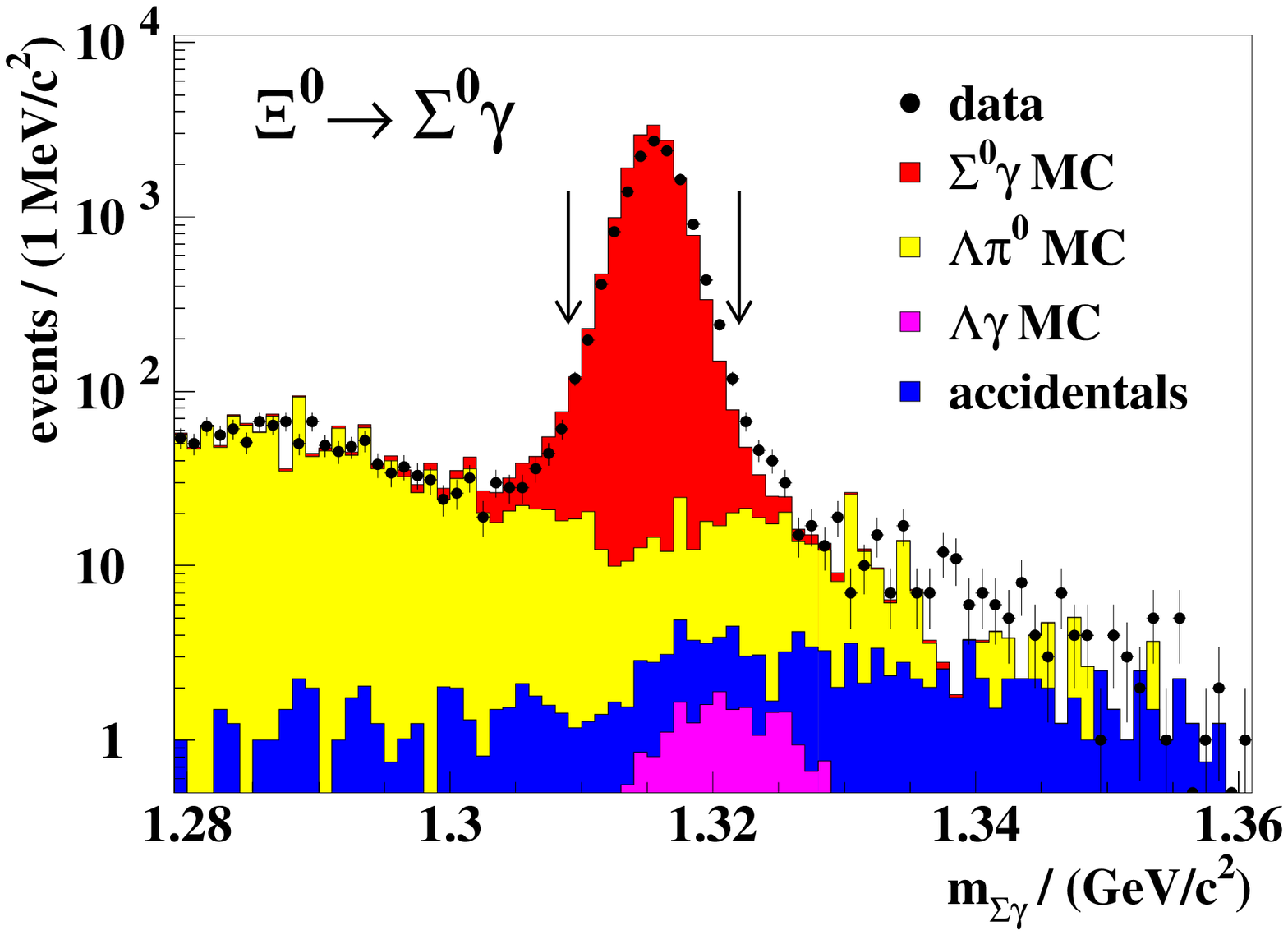,width=0.49\textwidth}
\caption{\label{fig:masses} $\Lambda \gamma$ (left) and $\Sigma^0 \gamma$ (right) invariant mass
         distributions of the selected signal events.
         Shown are also the different background contributions, as determined from
         the simulation.}
\end{center}
\end{figure}

With these criteria, 52~318 $\xilamgam$, 15~895 $\xisiggam$, and about 4~million $\xilampi$ candidates were selected.
The invariant mass distributions of the selected $\xilamgam$ and $\xisiggam$ events are shown in Fig.~\ref{fig:masses}.
The selection efficiencies were about
$3.4\%$ and $0.4\%$ for $\xilamgam$ and $\xisiggam$, respectively,
including the branching fraction of $\Lambda \to p \pi^-$.
In both channels, the main background are mis-identified $\xilampi$ events.
In $\xilamgam$ this background amounts to $0.6\%$, while in $\xisiggam$, due to
the similar signature, $\xilampi$ contributes to $1.5\%$ to the signal candidates.
Additional backgrounds from accidentally overlapping events or from cross-feed between the
two signal channels are at the $0.1\%$ level.
%The Monte Carlo simulation was not perfectly able to describe the tails of the mass distributions of the data.
%It was checked that this disagreement has no impact on the measured asymmetries.
Since the Monte Carlo simulation was not perfectly able to describe the tails of the mass distributions of the data,
the background from $\xilampi$ was estimated by fitting an exponential function to the side bands of the signal region.
The uncertainty from this estimation has only little impact on the measured asymmetries and is included in the systematic uncertainties.

\section{Monte Carlo Simulation}

A Monte Carlo simulation was performed with the full detector description 
based on the GEANT3 package~\cite{bib:geant}, including measured inefficiencies in the drift chambers, but without trigger simulation.
In total, 50~million $\xilamgam$, 100~million $\xisiggam$, and 300~million $\xilampi$ events
were generated.

The production plane was the $y$-$z$ plane; symmetry arguments imply that there was no $\Xi^0$ polarization in $y$ or $z$.
For the $\Xi^0$ polarization in $x$ direction, a value of $P_x = - 10\%$ was used, in agreement with
the measured value of $P_x^{\Xi^0} = (-9.7 \pm 0.7 \pm 1.3)\%$ for an unpolarized proton beam with energy of 800~GeV and a production
angle of 4.8~mrad~\cite{bib:ktevpol}, which covers the same $x_f$ region as in NA48/1.
For $\aXi$ hyperons, no polarization in the $x$ direction was assumed in agreement with the same KTeV measurement.
%For the decay asymmetry of the $\XiLampi$ and the subsequent $\Lamppi$ decay the world averages of
%%$\alpha(\Xi^0) \alpha_-(\Lambda) = -0.264 \pm 0.013$ was used~\cite{bib:pdg06}.
%$\alpha(\xilampi) = -0.411 \pm 0.022$ and $\alpha_-(\Lambda \to p \pi^-) = -0.642 \pm 0.013$ were used~\cite{bib:pdg06}.

The $\Xi^0$ beam profile was not perfectly simulated; it depends critically on precise know\-ledge
of the geometry of the target and collimator region and the beam.
The simulated beam profile was somewhat narrower than that found in the data.
This difference was taken into account by reweighting the simulated events with $r_\text{cog} > 3.75$~cm.

The $\Xi^0$ mass spectrum of the data appeared shifted up by about 0.5~MeV/$c^2$, or 2.5~standard deviations, 
with respect to the previously measured value~\cite{bib:xi0mass}.
We observed this shift for all $\Xi^0$ decay channels, but not for other
decaying particles, e.g.\ $\Sigma^0$ hyperons and neutral kaons. In the simulation we used this shifted value
of the $\Xi^0$ mass, and the discrepancy with the previously measured value
was taken into account in the systematic uncertainty of the asymmetry measurements.

\section{Data Analysis}

\subsection{Method of the Asymmetry Measurements}

\begin{figure}[t]
\begin{center}
\epsfig{file=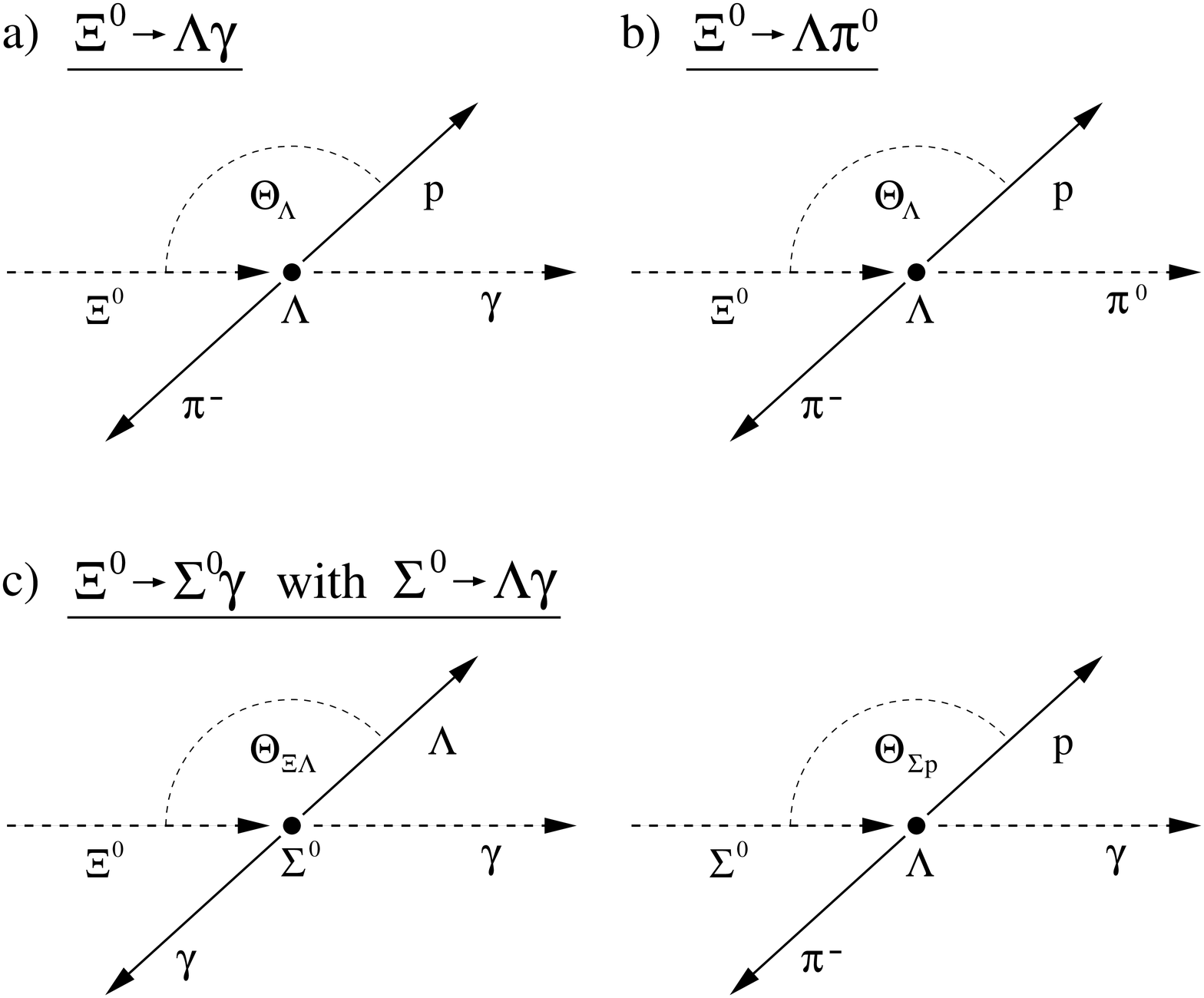,width=0.75\textwidth}
\caption{\label{fig:asym}Definition of the decay angles. 
         a) \& b) Decay angle $\Theta_\Lambda$ in the $\Lambda$ rest frame, as defined for $\xilamgam$ and $\xilampi$, respectively.
         c) Decay angles $\Theta_{\Xi \Lambda}$ and $\Theta_{\Sigma p}$ for the $\xisiggam$ and and the subsequent $\siglamgam$ decay,
            defined in the $\Sigma^0$ and $\Lambda$ rest frames, respectively.}
\end{center}
%\spaceafterfloat
\end{figure}

For the asymmetry measurements we exploited the well-known decay asymmetry of the $\lamppi$ decay.
For $\xilamgam$, the $\Lambda$ hyperons are longitudinally polarized by the parent process $\xilamgam$
with a mean polarization of $\alphalg$ in their rest frame.
Effectively, one measures the distribution of the angle $\Theta_\Lambda$ between the incoming $\Xi^0$ 
%(corresponding to the outgoing $\Lambda$ direction in the $\Xi^0$ rest frame) 
and the outgoing proton in the $\Lambda$ rest frame (see Figure \ref{fig:asym}):
\be
\frac{dN}{d\cos\Theta_\Lambda} \; = \; N_0 \left( 1 - \alphalg \, \alphapp \, \cos \Theta_\Lambda \right)
\label{eqn:lamgam}
\ee
In this way the $\Lambda$ is polarized by the $\Xi^0$ decay and analyzed by its own decay into $p \pi^-$.
The minus sign is purely conventional and arises from the fact that the photon carries spin~1, which leads to an opposite
$\Lambda$ spin to that in the process $\xilampi$~\cite{bib:behrends58}.

For $\xisiggam$, the situation is somewhat more complicated, as also the intermediate,
purely electro-magnetic decay $\siglamgam$ and the two decay angles 
$\Theta_{\Xi \Lambda}$ and $\Theta_{\Sigma p}$ of the $\Sigma^0$ and $\Lambda$ decays
have to be considered (see Fig.~\ref{fig:asym}~c)).
This leads to
\begin{eqnarray}
\frac{dN}{d \, (\cos\Theta_{\Xi \Lambda} \cos \Theta_{\Sigma p})} & = & N_0 \left( 1 + \alphasg \, \alphapp \, \cos\Theta_{\Xi \Lambda} \cos \Theta_{\Sigma p} \right) \nonumber \\
                                                              &   & \times \: F(\cos\Theta_{\Xi \Lambda} \cos \Theta_{\Sigma p})
\label{eqn:siggam}
\end{eqnarray}
where $F$ is a known function of $\cos\Theta_{\Xi \Lambda} \cos \Theta_{\Sigma p}$~\cite{bib:matthiasphd}.

For calibration purposes and as a cross-check, we also analyzed the decay $\xilampi$,
for which the decay asymmetry is well measured.
Obviously, when replacing $\alphalg$ with $\alphalp$,
there is no difference in the topology for the decays $\xilamgam$ and $\xilampi$
and the definition of the angle $\Theta_\Lambda$ is similar.
However, as explained above, the spin~0 nature of the $\pi^0$ 
leads to a sign flip for the longitudinal $\Lambda$ polarization.

\subsection{Trigger Efficiency Correction}

For the measurement of the decay asymmetries, the trigger efficiencies had to be taken into account.
It was determined separately for the different trigger levels 
and for different run periods, and as a function of either of the angular parameters
$\cos{\Theta_\Lambda}$ and $\cos\Theta_{\Xi \Lambda} \cos \Theta_{\Sigma p}$.

The L1 efficiency was measured using data taken with a minimum-bias trigger based on 
information from the hodoscope for neutral particles, 
which was independent from the trigger for radiative hyperon decays and down-scaled by a factor of 100.
From $\xilampi$ decays, an L1 efficiency of $99.8\%$ 
was obtained, independent of
the decay angle\footnote{Except for the first run period, where the L1 efficiency was between 85\% and 90\%.}.
For the L2 efficiency determination, L1 triggered data were used, which passed a minimum bias trigger for 
charged particles, down-scaled by 25 or 35, depending on the run period.
The obtained efficiencies range between 70\% and 95\%, depending on run period, decay channel, 
and decay angle. The efficiency correction was applied separately for each run period.
Finally, the efficiency of the L3 software trigger was determined to be larger than $99.9\%$
and was not further considered for the decay asymmetry measurement.

\subsection{Decay Asymmetry Measurements}

\begin{figure}[t]
\begin{center}
\epsfig{file=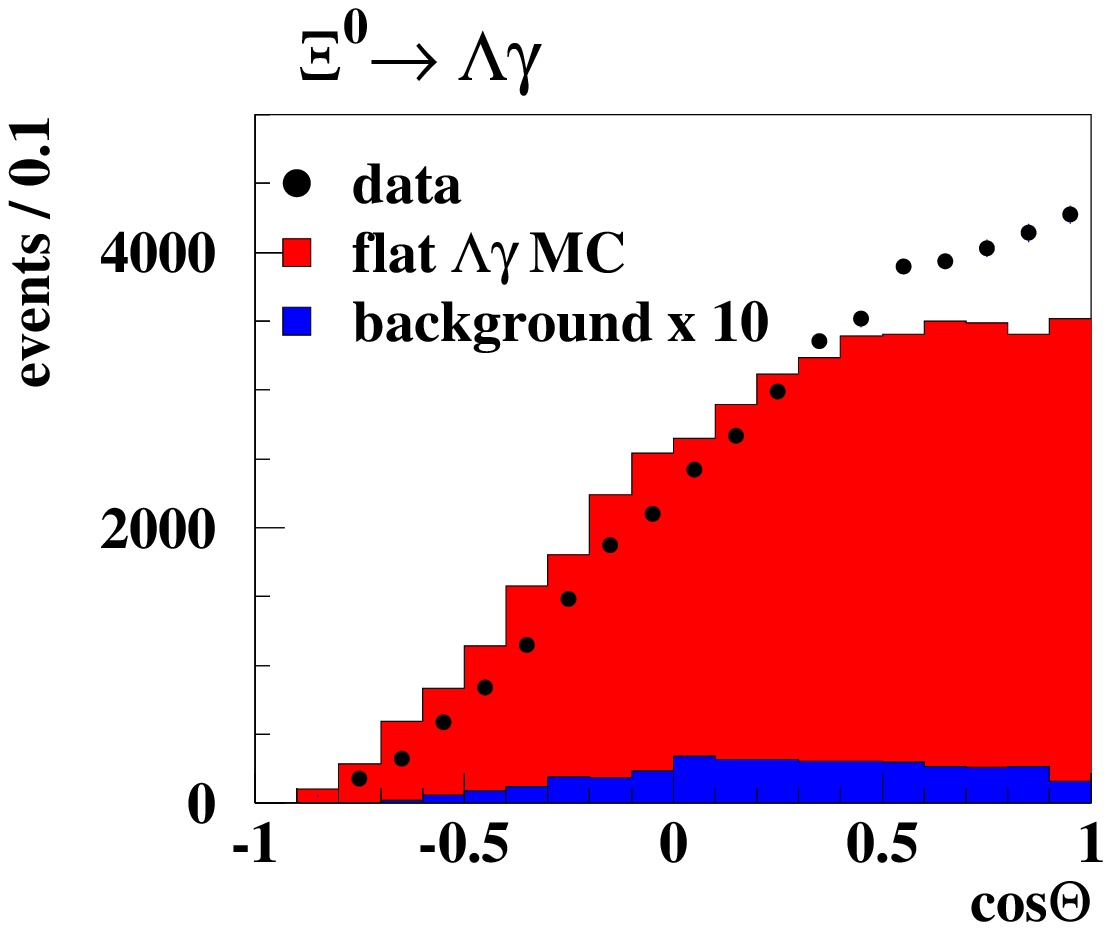,width=0.49\textwidth}
\epsfig{file=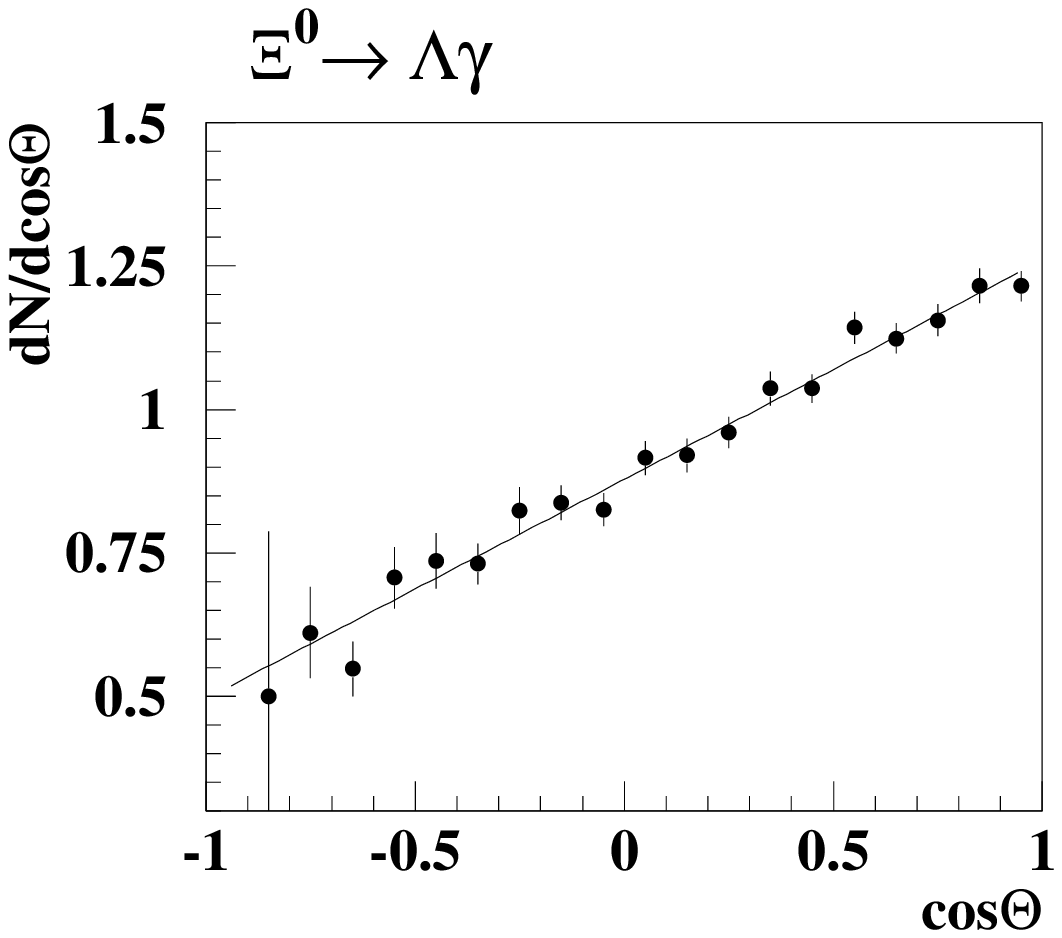,width=0.49\textwidth}
\caption{\label{fig:lamgamasym} 
         Left: $\xilamgam$ $\cos{\Theta_\Lambda}$ distributions for data, an isotropic and trigger-corrected MC simulation and simulated background events.
         Right: Ratio of background-corrected signal events over the MC simulation. The line shows the fit result.}
\end{center}
\end{figure}

The distribution of $\cos{\Theta_\Lambda}$ of all selected 
$\xilamgam$ events is shown in Fig.~\ref{fig:lamgamasym} (left)
together with an isotropic Monte Carlo simulation and simulated background events.
The simulated events were corrected for trigger efficiency.
The ratio of the background-subtracted data over Monte Carlo, shown in Figure~\ref{fig:lamgamasym} (right),
corrects for the detector acceptance and exhibits the expected linear slope.

The measurement of the decay asymmetry was performed with a least-squares fit which compared the data events
with the trigger-efficiency and background corrected Monte Carlo simulation. Free fit parameters were 
the product of decay asymmetries $\alphalg \, \alphapp$ and the overall normalization $N$.
To be insensitive to resolution effects (which might originate from not using the true, but the reconstructed $\cos{\Theta_\Lambda}$ variable),
in the fit a Monte Carlo with a decay asymmetry close to the measured one was used. In this way, the fit effectively measured 
the ratio $N \, (1- \alphalg \, \alphapp \, \cos{\Theta_\Lambda})/(1- \alphalg^\text{MC} \, \alphapp^\text{MC} \, \cos{\Theta_\Lambda})$,
which is close to a constant.
The data were fitted in the range $-0.8 < \cos{\Theta_\Lambda} < 1.0$, where each bin had at least 20 data and 40 Monte Carlo entries.
As fit result we obtained $\alphalg \, \alphapp = -0.452 \pm 0.012_\text{stat}$, corresponding to
$\alphalg = -0.704 \pm 0.019_\text{stat}$, with the uncertainty from the signal statistics.

\begin{figure}[t]
\begin{center}
\epsfig{file=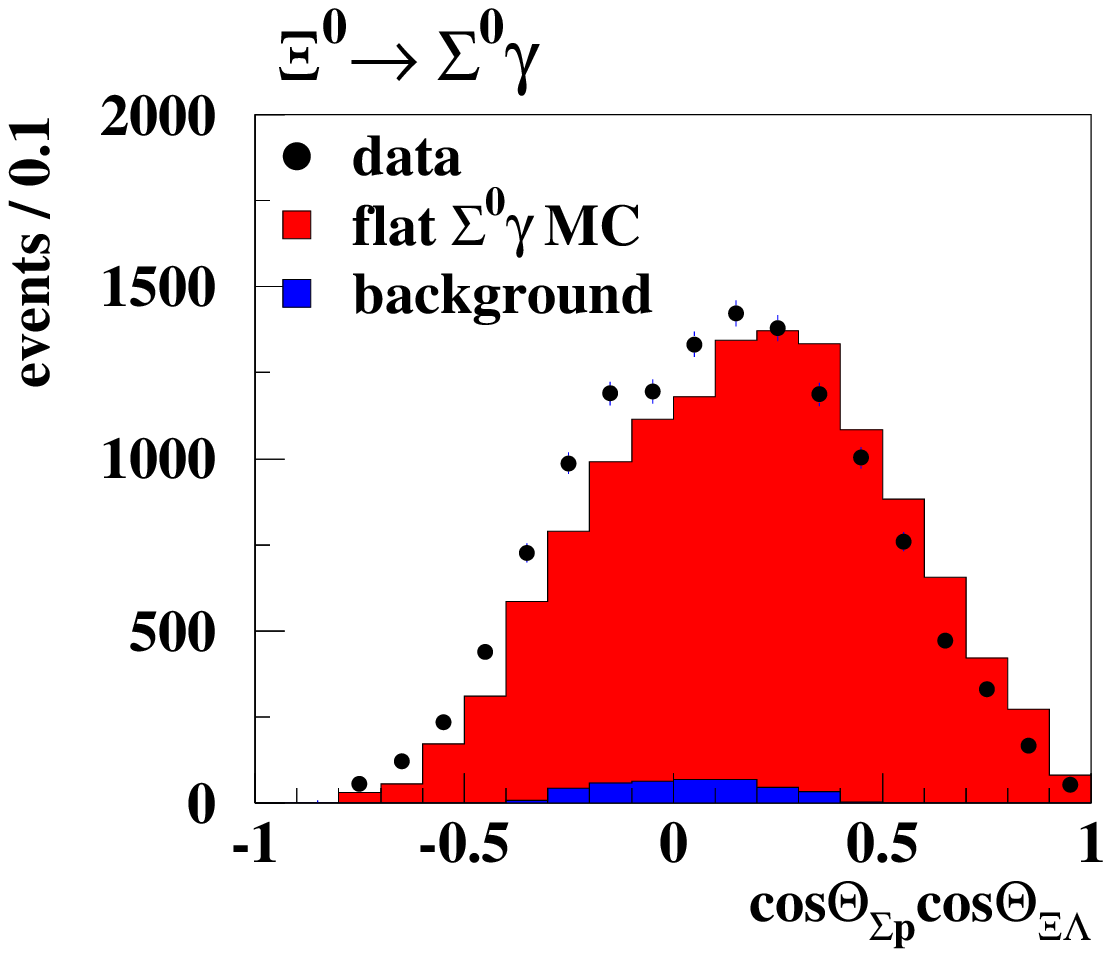,width=0.49\textwidth}
\epsfig{file=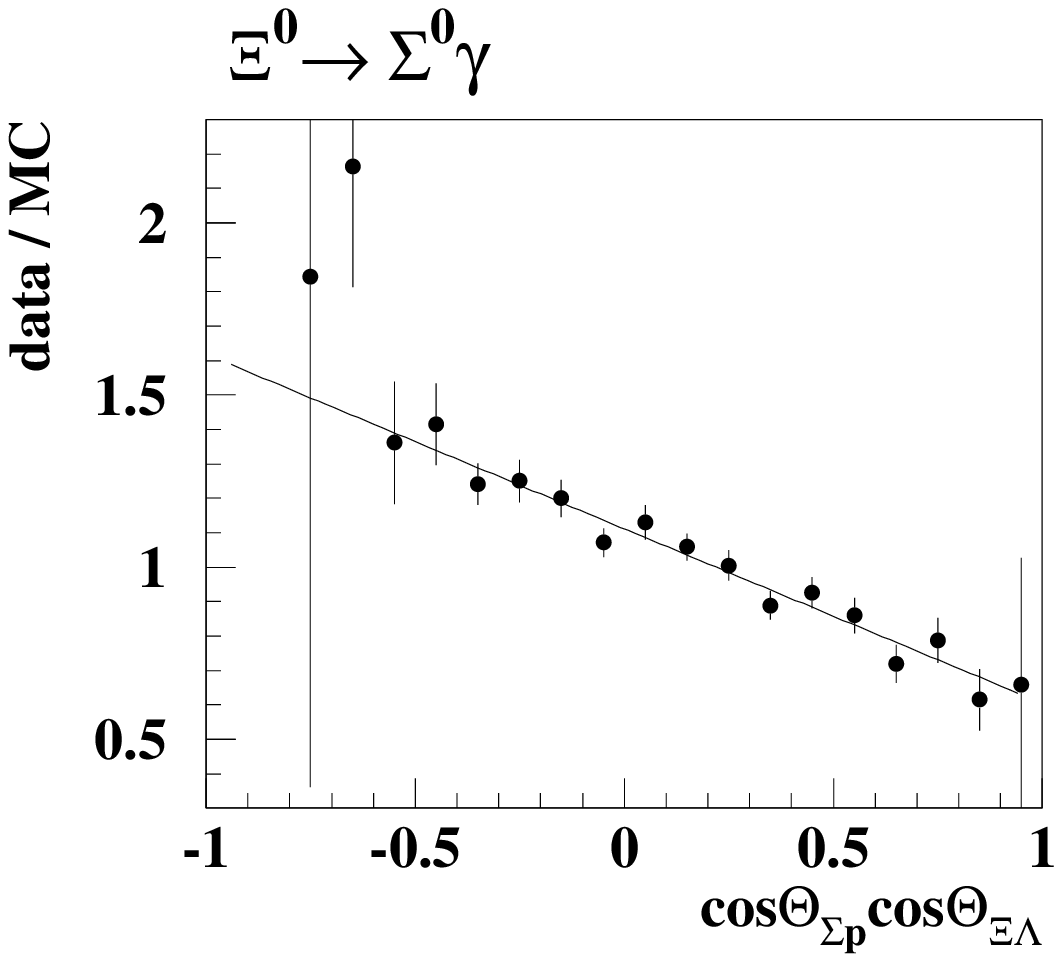,width=0.49\textwidth}
\caption{\label{fig:siggamasym} 
         Left: $\xisiggam$ distributions of $\cos\Theta_{\Xi \Lambda} \cos \Theta_{\Sigma p}$ for data, an isotropic MC simulation and simulated background events.
         Right: Ratio of background-corrected signal events over the isotropic and trigger-corrected MC simulation. The line shows the fit result.}
\end{center}
\end{figure}

The angular distributions of $\cos\Theta_{\Xi \Lambda} \cos \Theta_{\Sigma p}$ for
$\xisiggam$ are shown in  Fig.~\ref{fig:siggamasym}
together with an isotropic Monte Carlo simulation and simulated background events.
The simulated events were corrected for trigger efficiency.
The fit was performed completely analogous to $\xilamgam$ using a least-squares fit and a Monte Carlo simulation 
with a decay asymmetry close to the measured one.
The fit range was $-0.6 < \cos\Theta_{\Xi \Lambda} \cos \Theta_{\Sigma p} < 0.9$.
The fit result is $\alphasg \, \alphapp = -0.468 \pm 0.020_\text{stat}$, or $\alphasg = -0.729 \pm 0.030_\text{stat}$,
where the error again only contains signal statistics.
Despite the same sign of $\alphasg$ and $\alphalg$, the sign of the slope in Fig.~\ref{fig:siggamasym}
is opposite to the slope in $\xilamgam$ due to the spin of the additional photon.

\subsection{Systematic Uncertainties}

Several contributions to the systematic error were studied.

Trigger efficiencies were measured from the data as described above and contributed
uncertainties of $\pm 0.020$ and $\pm 0.028$ to $\alphalg$ and $\alphasg$, respectively.

The largest uncertainty came from a possible incorrect modeling of the detector acceptance.
The critical point was the simulation of the beam profile in the beam pipe region, since the acceptance changed rapidly 
with the distance of the decay from the beam axis.
This was investigated with $\xilampi$ data events, which have sufficient statistics, and 
which were, due to the lower $Q$ value, more sensitive to possible inefficiencies near the beam pipe.
Observed variations of the fitted $\xilampi$ decay asymmetry with the $\Xi^0$ energy 
and with the minimum distance between the photon clusters and the track impact points in the calorimeter were applied on the
radiative decays, resulting in variations 
of $\pm 0.058$ and $\pm 0.067$ on the $\xilamgam$ and $\xisiggam$ decay asymmetries, respectively.
Even though the acceptance variations are less strong for $\xilamgam$ and $\xisiggam$, these values were taken 
as a conservative estimate of the corresponding systematic uncertainty.

The systematic uncertainty arising from background subtraction was conservatively estimated by varying the background
in $\xilamgam$ by $\pm 25\%$ and in $\xisiggam$, where the remaining contribution of $\xilampi$ events 
was more difficult to estimate, by $\pm 100\%$. 

The uncertainty of our knowledge of the $\Xi^0$ mass was determined by shifting the $\Xi^0$ mass
in the simulation to the PDG value.
Finally, the known uncertainties of the $\Xi^0$ lifetime~\cite{bib:pdg08}, the transverse $\Xi^0$ polarization,
and of the PDG value of $\alphapp$ were taken into account.

The single contributions to the systematic uncertainty are summarized in Table~\ref{tab:systematics}.
For the total systematic uncertainty, the single components were added in quadrature.

\begin{table}[th]
\begin{center}
\begin{tabular}{lrr}
\hline \hline
                                     & $\Delta \: \alphalg$ & $\Delta \: \alphasg$ \\ \hline
Trigger efficiencies                 &   $\pm \: 0.020$     &   $\pm \: 0.028$     \\
Detector acceptance                  &   $\pm \: 0.058$     &   $\pm \: 0.067$     \\
Background                           &   $\pm \: 0.003$     &   $\pm \: 0.008$     \\
$\Xi^0$ mass                         &   $\pm \: 0.009$     &   $\pm \: 0.009$     \\ 
$\Xi^0$ lifetime                     &   $\pm \: 0.002$     &   $\pm \: 0.002$     \\ 
$\Xi^0$ polarization                 &   $\pm \: 0.003$     &   $\pm \: 0.013$     \\ 
$\alphapp = 0.642 \pm 0.013$         &   $\pm \: 0.014$     &   $\pm \: 0.015$     \\ \hline 
Total systematic uncertainty         &   $\pm \: 0.064$     &   $\pm \: 0.076$    \\*[1mm]
Statistical uncertainty              &   $\pm \: 0.019$     &   $\pm \: 0.030$    \\
\hline \hline
\end{tabular}
\caption{Summary of uncertainties on the measured $\xilamgam$ and $\xisiggam$ decay asymmetries.}
\label{tab:systematics}
\end{center}
%\spaceafterfloat
\end{table}

An important check for the validity of the data analysis was the measurement of the decay asymmetry in $\xilampi$, where
the data statistics is a factor of 100 higher.
% All channels are two-body decays and the decay $\xilampi$, due to the lower $Q$ value,
% is even more sensitive to effects from the beam-line geometry and polarization than the radiative decays.
The decays $\xilampi$ were recorded with the same trigger and used a similar Monte Carlo simulation
as for the other analyses.
From our data we measured a combined $\xilampi$ asymmetry of $\alphalp \alphapp = -0.276 \pm 0.001_\text{stat} \pm 0.035_\text{syst}$.
%The systematics of this measurement were evaluated with the same method as for the main analyses. 
The systematics of this measurement are mainly due to a dependency of the fitted asymmetry with the $\Xi^0$ energy and
were estimated from the variations of the result for different $\Xi^0$ energies.
The agreement between this result and the best published measurement of $-0.260 \pm 0.006$~\cite{bib:handler82}
validated the method and served as a systematic check.

As a further check, an independent analysis, which was performed using the same data but different Monte Carlo samples,
gave consistent results.

\subsection{Measurement of $\aXi$ Decay Asymmetries}

In addition to $\Xi^0$ decays, we have also selected $\axilamgam$ and $\axisiggam$
events. The selection criteria were identical, but opposite track charges were required.
In total, we found 4769 $\axilamgam$ and 1404 $\axisiggam$ candidates in the same data set.
Backgrounds were estimated to be $1.7\%$ for $\axilamgam$, mainly arising from accidental overlaps,
and $1.5\%$ for $\axisiggam$, dominated by $\axilampi$ events.

The decay asymmetry of these channels were determined in exactly the same way as for the
$\Xi^0$ decays, but, due to the limited statistics, with somewhat smaller fit regions in the 
angular variables and using a maximum-likelihood method instead of a least-squares fit.
The results were 
$\alphaalg = -0.798 \pm 0.064_\text{stat}$ and
$\alphaasg = -0.786 \pm 0.104_\text{stat}$, in agreement with the $\Xi^0$ decay asymmetries,
but with much larger statistical uncertainties.

\section{Conclusions}

From the data of the NA48/1 experiment, we found the $\xilamgam$ and $\xisiggam$ decay asymmetries to be
\be 
\alphalg = -0.704 \pm 0.019_\text{stat} \pm 0.064_\text{syst}
\ee
and
\be 
\alphasg = -0.729 \pm 0.030_\text{stat} \pm 0.076_\text{syst}.
\ee

The first result is a factor of three more precise than the previous measurement
by NA48~\cite{bib:lamgamNA48} while the latter is of similar accuracy as the KTeV measurement~\cite{bib:siggamKTeV},
and both are in very good agreement with the previous measurements.

With these measurements large negative decay asymmetries have been established in both channels. 
Unfortunately the theoretical uncertainties are still rather large, in particular those of the $\xilamgam$ channel.  
We hope that this situation will improve in the future.

\section{Acknowledgements}

It is a pleasure to thank the technical staff of the participating
laboratories, universities, and affiliated computing centres for their
efforts in the construction of the NA48 apparatus, in the
operation of the experiment, and in the processing of the data.

%
%  Bibliography
%

\end{document}